\documentclass[groupedaddress,jap,aip,preprint,graphicx]{revtex4-1} 
 \pdfoutput=1
\usepackage{graphicx}
\usepackage{epstopdf}
\usepackage{times}
\usepackage[small, compact]{titlesec}
\begin{document}
\title{Disappearance of Electron-Hole Asymmetry in Nanoparticles of $\bf Nd_{1-x}Ca_xMnO_3$ $\bf (x = 0.6, 0.4)$: Magnetization and Electron Paramagnetic Resonance Evidence} 

\author{Bhagyashree K. S.} 
\author{S. V. Bhat}
\affiliation{Department of Physics, Indian Institute of Science, Bangalore-560012, India}

\begin{abstract}
  We study and compare magnetic and electron paramagnetic resonance behaviors of bulk and nanoparticles of $Nd_{1-x}Ca_xMnO_3$ in hole doped $(x = 0.4; NCMOH)$ and electron doped $(x = 0.6; NCMOE)$ samples. NCMOH in bulk form shows a complex temperature dependence of magnetization M(T), with a charge ordering (CO) transition at $\sim$ 250 K, an antiferromagnetic (AFM) transition at $\sim$ 150 K and a transition to a canted AFM phase/mixed phase at $\sim$ 80 K. Bulk NCMOE behaves quite differently with just a charge ordering transition at  $\sim$ 280 K, thus providing a striking example of the so called electron-hole asymmetry. While our magnetization data on bulk samples are consistent with the earlier reports, the new results on the nanoparticles bring out drastic effects of size reduction. They show that M(T) behaviors of the two nanosamples are essentially similar in addition to the absence of the charge order in them thus providing strong evidence for vanishing of the electron-hole asymmetry in nanomanganites. This conclusion is further corroborated by electron paramagnetic resonance studies which show that the large difference in the 'g' values and their temperature dependences found for the two bulk samples disappears as they approach a common behavior in the corresponding nanosamples. 
\end{abstract}

\maketitle
\section{Introduction}
Doped perovskite manganites with the general formula $RE_{1-x}A_xMnO_3$, where ‘RE’ is a trivalent rare
earth ion and ‘A’ is a divalent alkaline earth ion display complex phase diagrams as a function of composition and temperature. The plethora of structural, magnetic and trasport phases exhibited by them involve charge ordering, orbital ordering, para, antiferro and ferromagnetism and metallic and insulating behaviors. The strong correlation between spin, lattice and orbital degrees of freedom present in these materials lead to a delicate balance between different energy scales leading to similar probabilities of occurrence of different phases~\cite{TokuraBook,Dagotto,Fath,Loudon,Tokura1,CNRrao}. This complex behavior is continuing to be a challenge to understand theoretically while at the same time being attractive from the point of view of applications because of the interesting phenomena such as colossal magnetoresistance.

The phase diagrams of the manganites also exhibit a striking asymmetry across half doping,  i.e. $x=0.5$. For example, in $Nd_{1-x}Ca_xMnO_3$ system, bulk NCMOH with $x = 0.4$ shows CO, AFM and canted AFM/mixed phases, whereas bulk NCMOE with $x = 0.6$ does not show any of these phases other than the CO phase~\cite{Daniel1,Zhang1}. This kind of
asymmetry is termed electron-hole asymmetry since for $0<x<0.5$, more and more holes are introduced in $REMnO_3$
by the substitution of $Mn^{3+}$ by $Mn^{4+}$ and for $0.5<x<1$, $Mn^{4+}$ in $AMnO_3$ are replaced by $Mn^{3+}$, effectively doping
with more and more electrons. It is by now well established that when the particles of bulk polycrystalline manganites of $\ge$ micron size are reduced to partciles of $< 100$ nm, the properties change drastically, owing to quantum confinement effects and large surface to volume ratio. 
Rao et al. reported that by reducing the size of half doped $Nd_{0.5}Ca_{0.5}MnO_3$ particles to 20nm the CO and AFM phases observed in the bulk vanish giving rise to FM phase~\cite{RaoNCMO}.
The objective of the present work is to prepare $<100$ nm sized nanoparticles of electron and hole doped NCMO and study the effect of size reduction on the electron-hole asymmetry observed in the bulk samples.
Our magnetization studies show significant changes in the properties of nanosamples compared to their bulk counterparts. We observe that in the nanosamples the CO phase has weakened and an FM phase has emerged  and more importantly the electron hole asymmetry observed in the bulk has practically disappeared.

\section{Preparation}\label{Preparation}
NCMOH and NCMOE nanoparticles were prepared by citrate-gel method. The stoichiometric ratios of $Nd_2O_3$, $CaCO_3$ and $MnCO_3$ are weighed with preheating of
$Nd_2O_3$ at 900 $^0C$ for 9 hours to remove moisture content if any and weighing within 15 minutes of taking out of the furnace. The weighed chemicals are converted into nitrates by adding nitric acid. Required  amount of citric acid was added and the solution was slowly evaporated on a hotplate to get a gel. The gel is then decomposed at 250 $^0C$ to get a dark brown powder, which was then sintered at 600 $^0C$ for 6 hours to get the nanoparticle samples. The bulk samples were prepared by first heating the precursor nanoparticles at 1000 $^0C$ for 12 hours and then by grinding and 
pelletizing to sinter at 1300 $^0C$ for 24 hours.
 
\section{Characterization}\label{Characterization}
\begin{figure}
\includegraphics[width=\linewidth]{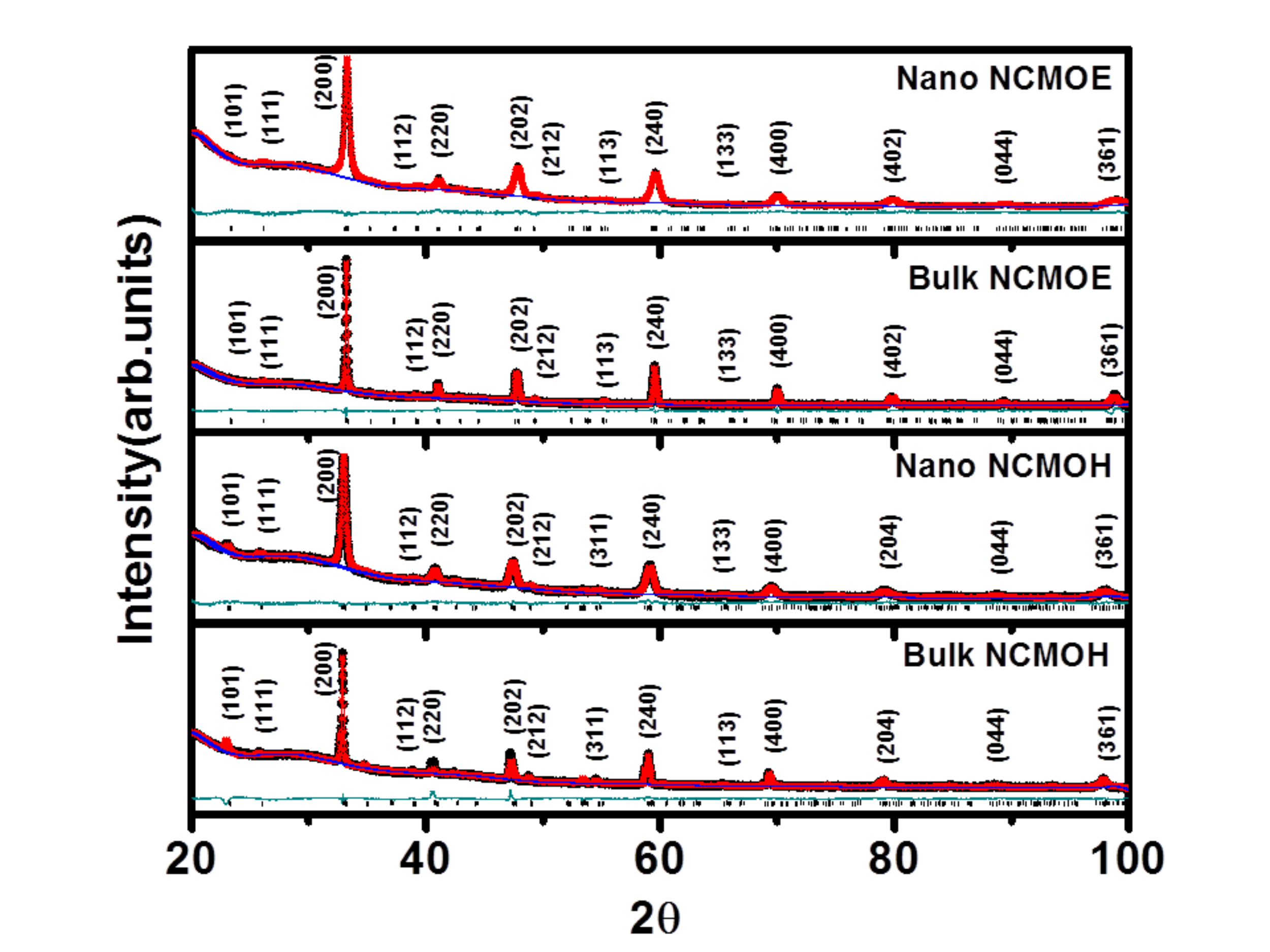}
\caption{ Rietveld refinement of XRD data of all the four samples, fitted to orthorhombic structure with space group Pnma}
\label{Xrd}
\end{figure}

 The samples were extensively characterized to check for purity and phase formation. 
X-ray diffraction was done in order to confirm the phase formation. Fig.\ref{Xrd} shows the Rietveld refined XRD results.
All the four samples crystallize in the orthorhombic structure with the space group Pnma.
For nano NCMOE, a = 5.4119, b = 7.5930,  c = 5.3731 \AA; $\chi^2$ = 1.692 and for bulk NCMOE, a = 5.4000, b = 7.5970, c = 5.3835 \AA;  $\chi^2$ = 3.973.
For nano NCMOH, a = 5.4546, b = 7.6431, c = 5.4158 \AA; $\chi^2$ = 1.881 and for bulk NCMOH, a = 5.4367, b = 7.6307, c = 5.3988 \AA; $\chi^2$ = 7.546.

\begin{figure}
\includegraphics[width=\linewidth, clip=true, trim=0cm 3.5cm 0cm 2cm]{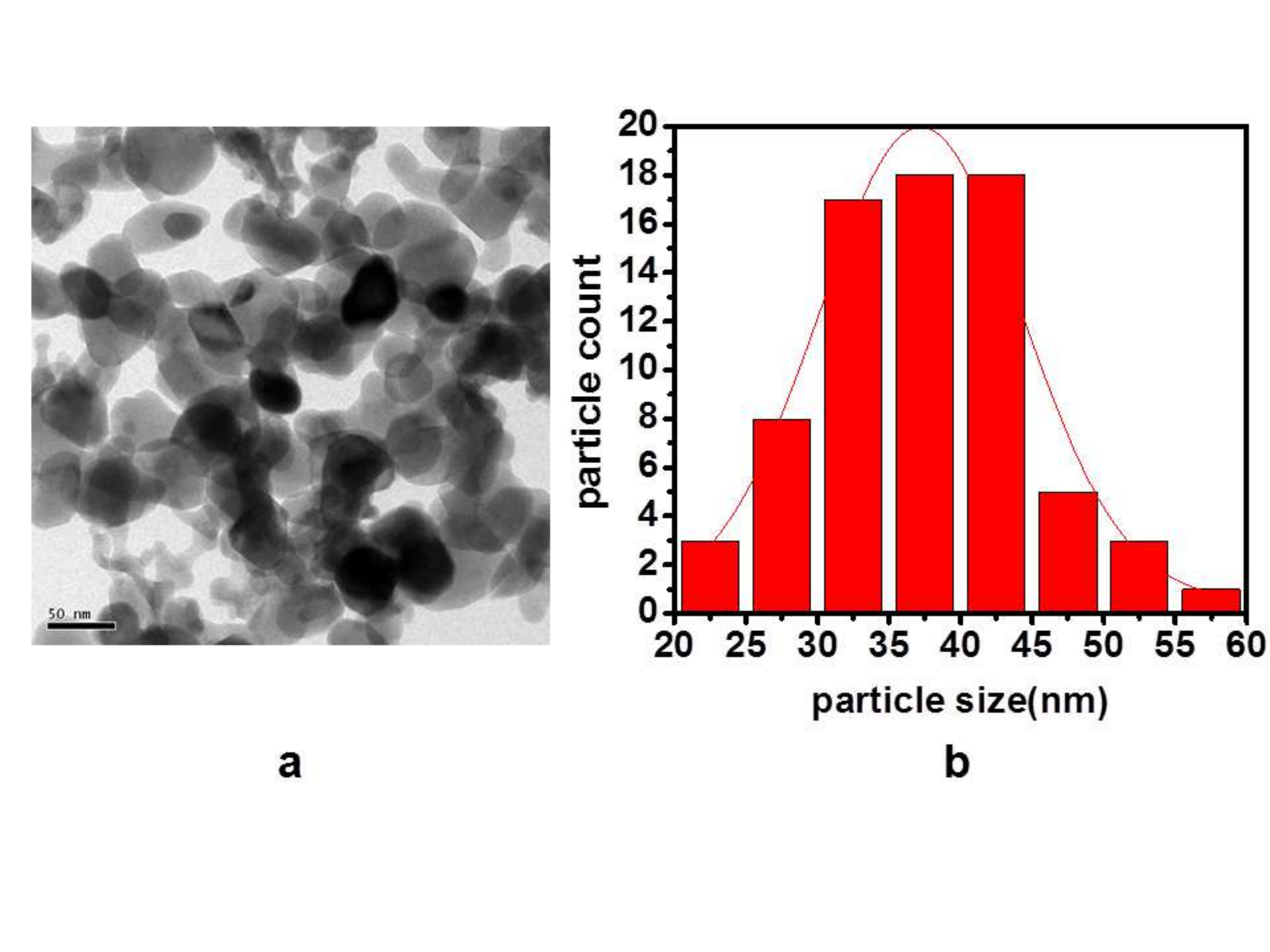}
\caption{ a)Nanoparticles of NCMOH. b)Histogram of NCMOH nanoparticles}
\label{TEM}
\end{figure}

The EDAX(Energy Dispersive X-ray Analysis) done on the samples indicated compositions very close to the intended stoichiometric ratios. Iodometric titration was done on the samples 
to calculate the oxygen stoichiometry, the average value of which was found to be $\sim3$. 
TEM(Transmission Electron Microscopy) was done on the two nano samples in order to find the particle size; the average particle size is found to be $\sim 40nm$ for both the samples. Fig.\ref{TEM}a shows the TEM picture of NCMOH nanoparticles. The figure shows that the particles are nearly mono-dispersed as substantiated by the histogram shown in Fig.\ref{TEM}b. 
Most particles have sizes in the range 30--45 nm. NCMOE nanoparticles also show similar size distribution.     

\section{Magnetization Studies}\label{Magnetization Studies}
\begin{figure}
\includegraphics[width=\linewidth]{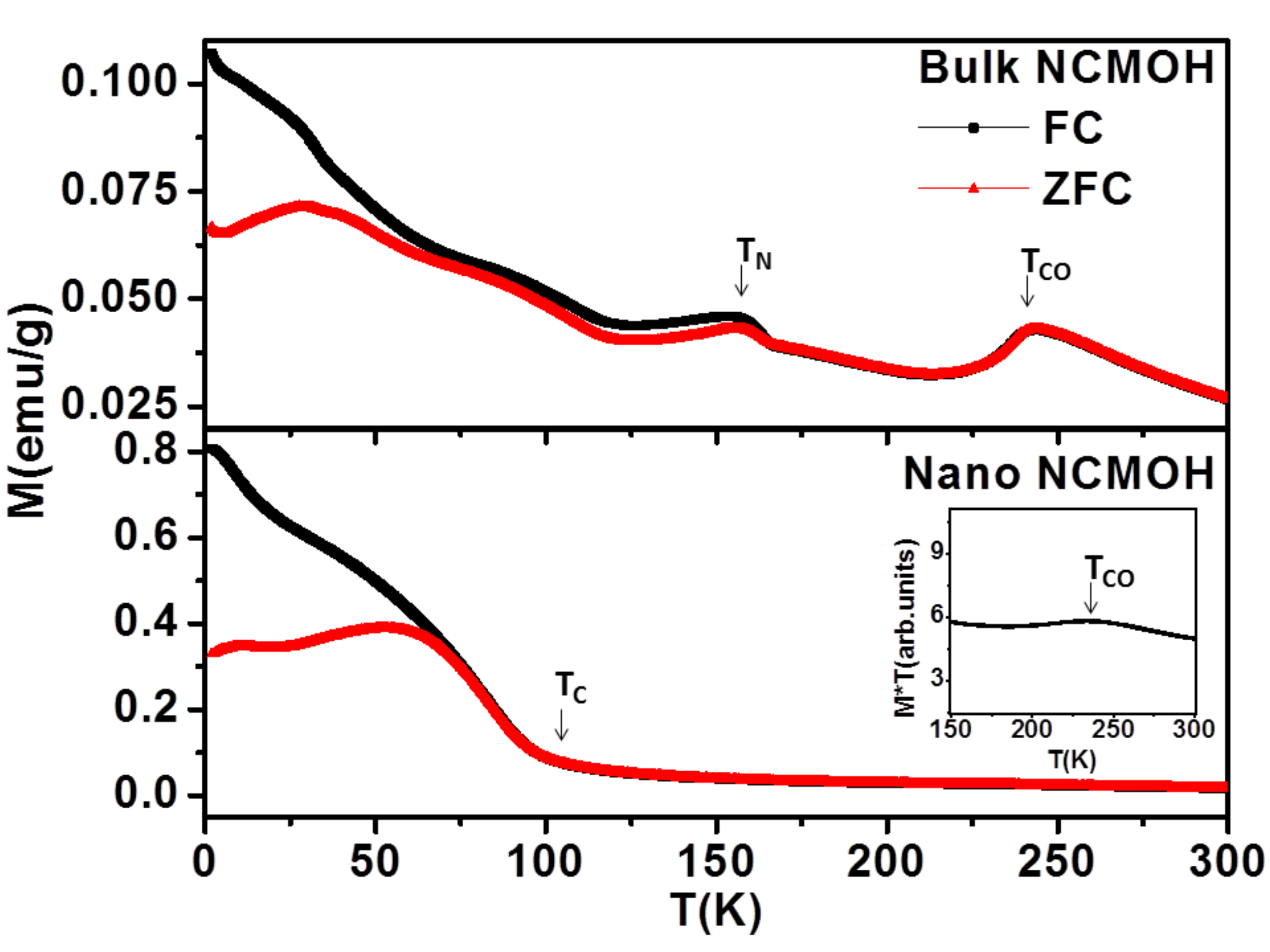}
\caption{ FC and ZFC magnetization behaviors of NCMOH bulk and nano samples }
\label{FZh}
\end{figure}

\begin{figure}
\includegraphics[width=\linewidth]{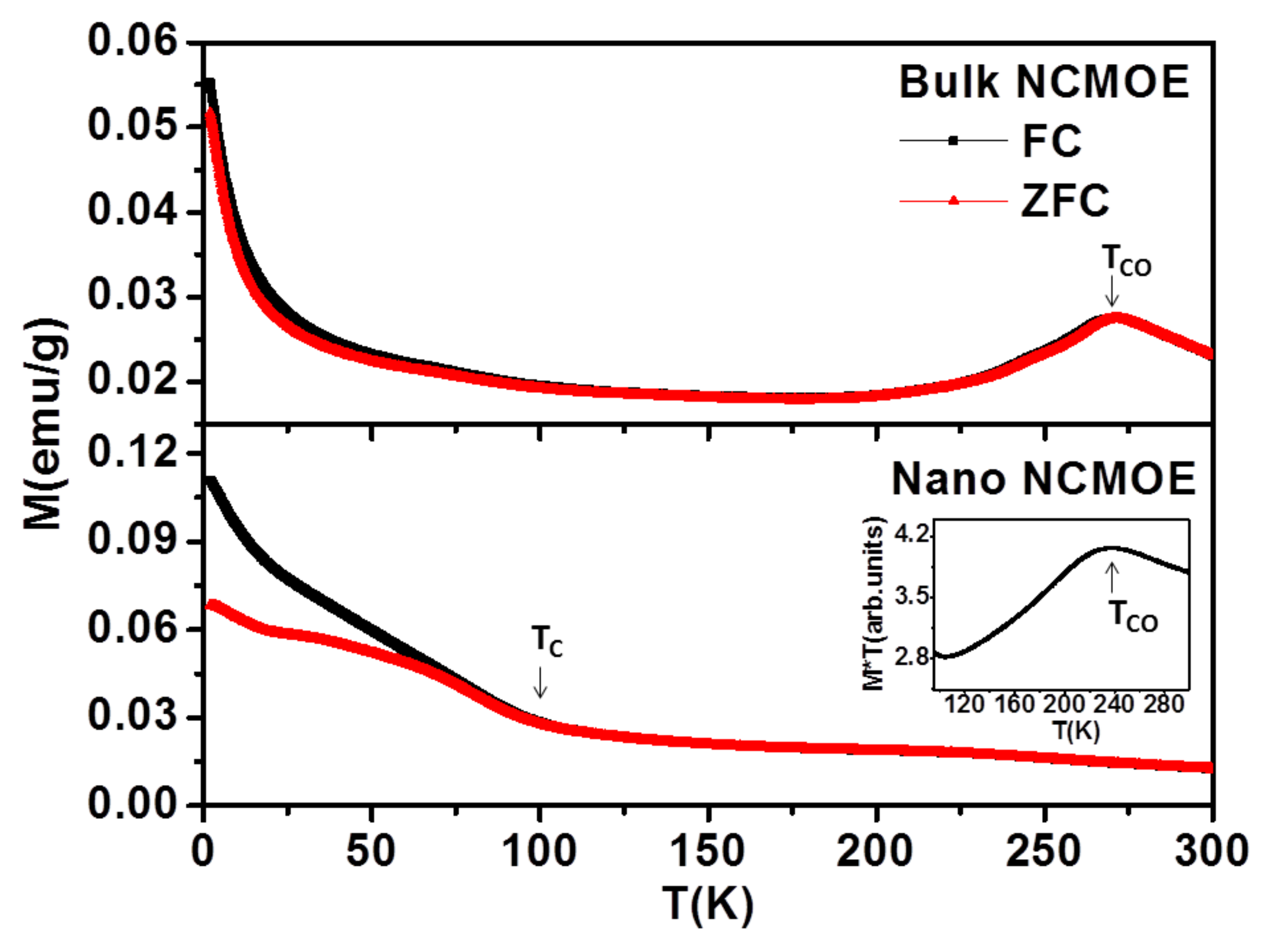}
\caption{ FC and ZFC magnetizations of NCMOE bulk and nano samples }
\label{FZe}
\end{figure}

The magnetization measurements were carried out using a commercial SQUID magnetometer. Field cooled (FC) under 100 Oe and zero field cooled (ZFC) magnetizations of the four samples were recorded in the temperature range from 2 K to 300 K under a measuring fleld of 100 Oe.
Fig.\ref{FZh} shows the plot of FC and ZFC magnetizations for bulk and naoparticles of NCMOH samples. The M(T) for the bulk sample exhibits a number of peaks signifying different magnetic transitions as the sample is cooled; a CO transition at $\sim$ 250 K which is evident by a sharp peak in magnetization curve due to localization of charge carriers and a weaker peak at $\sim$ 150 K indicating an AFM transition in accordance with the earlier report~\cite{Daniel1}. As the temperature is lowered further, the magnetization starts increasing at $\sim$ 120 K with a broad peak at $\sim$ 80 K which is attributed to the transition to a canted AFM phase in the earlier report. This peak could also be due to a FM transition leading to the existence of a mixed phase at lower temperatures. It is not possible to distinguish between a canted AFM phase and a mixed phase with magnetization measurements alone.
FC and ZFC M(T) behaviors of  nanoparticles of NCMOH are shown in the lower panel of Fig.\ref{FZh} and the contrast from the bulk behavior is striking. The multitude of peaks present in the bulk sample are conspicuous by their absence indicating that the CO, AFM and canted AFM/mixed phases have all disappeared and at $\sim 100$ K an FM like phase has emerged. In the inset to the figure, $M\times T$ vs $T$ is plotted to check if there is any residual CO. The curve shows a very weak and broad peak at $\sim$ 250 K pointing to the presence of short range CO in the nanoparticles~\cite{Zhou}. Fig.\ref{FZe} shows the FC and ZFC plots of NCMOE bulk and nanoparticles. The only transition that the bulk sample undergoes is a CO transition at $\sim$ 280 K, in accordance with the earlier report~\cite{Zhang1}. The nano NCMOE sample does not show any CO transition in the magnetization curve. It undergoes a transition to an FM phase at $\sim$ 100 K like the hole doped nanosample.   The inset of the figure shows $M\times T$ vs $T$ plotted in order to check if there is any residual CO present in the electron doped nanosample. A broad peak at $\sim$ 240 K is found which shows the presence of short range CO~\cite{Zhou}. The existence of residual CO in the two nanosamples could also be due to the presence of a small amount of polydispersity in the samples or due to the still moderately large diameter ($\sim 40$ nm) of the nanoparticles.
The blocking temperature $T_b$ in FC and ZFC curves of bulk NCMOH, nano NCMOH and nano NCMOE is $\sim$ 65 K. Below 50 K, there is a weak ferromagnetism observed in all the samples with an increase in magnetization in the FC curve, owing to the weak coupling of Nd ions with the Mn ions~\cite{Dupont}.                        

\begin{figure}
\includegraphics[width=\linewidth]{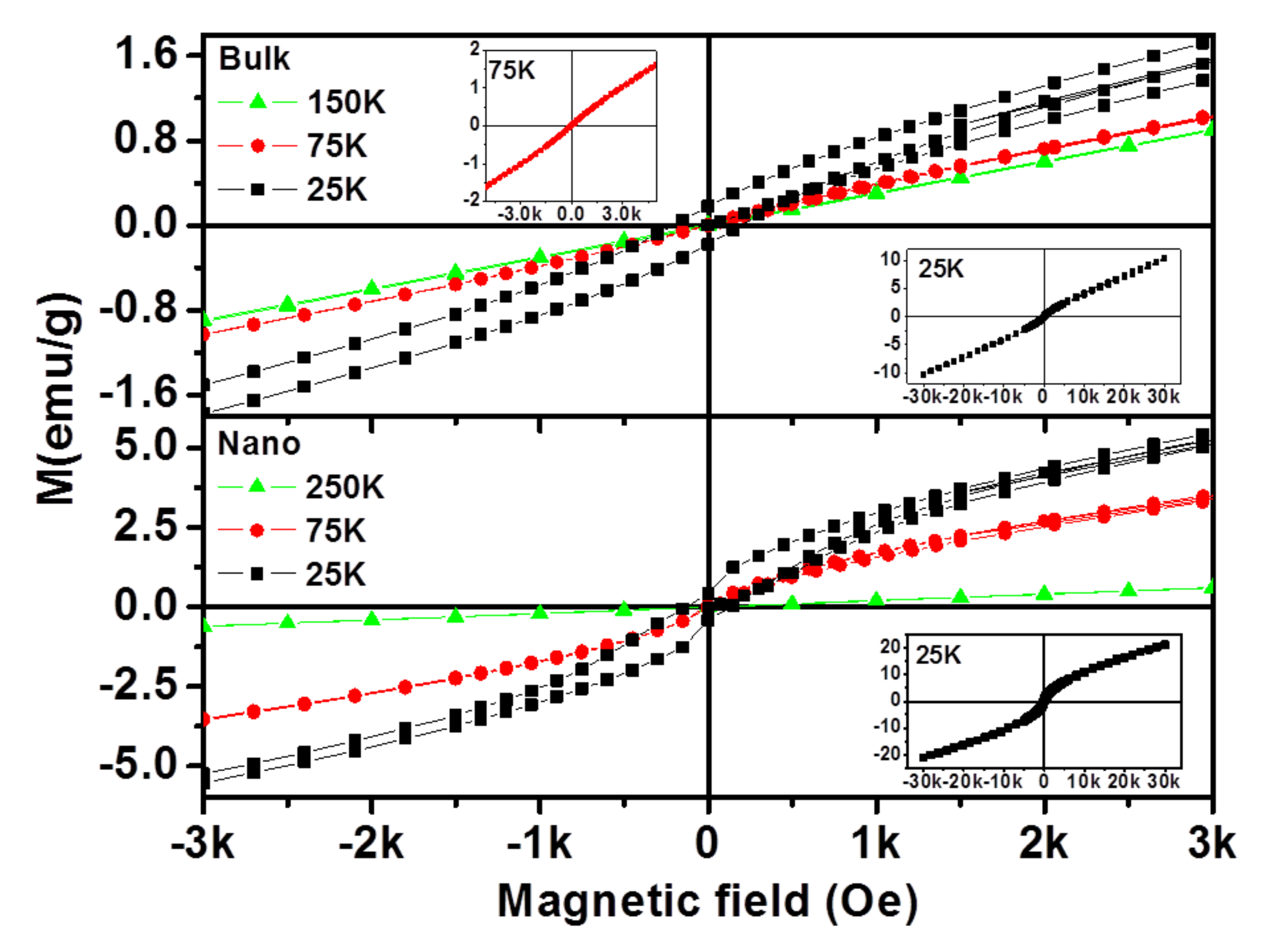}
\caption{ M vs H of NCMOH bulk and nano samples }
\label{MHh}
\end{figure}

M vs H measurements were performed on all the samples at different temperatures up to 3 T field. Fig.\ref{MHh} shows the results for the bulk and nano hole doped samples. The upper part of the Fig.\ref{MHh} shows the field dependent magnetization at 150 K, 75 K and 25 K for bulk sample. At 150 K the sample is in the AFM phase and the curve is linear as expected. At 75 K which is near the transition the curve is not completely linear possibly indicating the presence of more than one phase. At 25 K we observe hysteresis in the curve. The M(H) curve does not get saturated (inset of figure) even at the highest field applied. We attribute this to the presence of either a canted AFM phase or a mixed phase.
The lower part of Fig.\ref{MHh} shows the M vs H of nanosamples at 250 K, 75 K and 25 K. At 250 K the sample is in PM phase and shows a linear curve. At 75 K the curve is clearly non linear as the sample is in the FM phase. At 25 K the curve shows hysteresis indicating the presence of FM phase. The M(H) curve does not saturate (inset of the figure) till the highest field  
which might be due to the presence of an antiferromagnetic phase as proposed in the core shell model~\cite{Zhang4,Zhang3,Dong1,Dong2} of the nanoparticles.
According to this model the nanoparticle consists of an antiferromagnetic core surrounded by a ferromagnetic shell.   
Fig.~\ref{MHe} shows the M vs H plot for bulk and nano electron doped samples. The upper part of Fig.~\ref{MHe} shows the field dependent magnetization at 280 K and 25 K. The dependence is linear at both the temperatures. This indicates that the sample is in a single phase even at 25 K. 
The lower part of Fig.~\ref{MHe} shows the M vs H for the nanosample measured at 220 K, 75 K and 25 K. At 220 K the plot is linear as expected for a paramagnetic phase. At 75 K and 25 K i.e., in the FM phase hysteresis is observed but the curves do not saturate even at the maximum field that was applied. This observation can also be explained in terms of an antiferromagnetic core and a ferromagnetic shell as in the case of nano NCMOH. Exchange bias (EB) effects arising due to the coupling between these AFM and FM regions are expected and indeed observed in NCMOH and NCMOE as in many other nanomanganite systems. The detailed studies on EB as well as AC susceptibility results on the ground states of the present nanosamples will be published elsewhere.

\begin{figure}
\includegraphics[width=\linewidth]{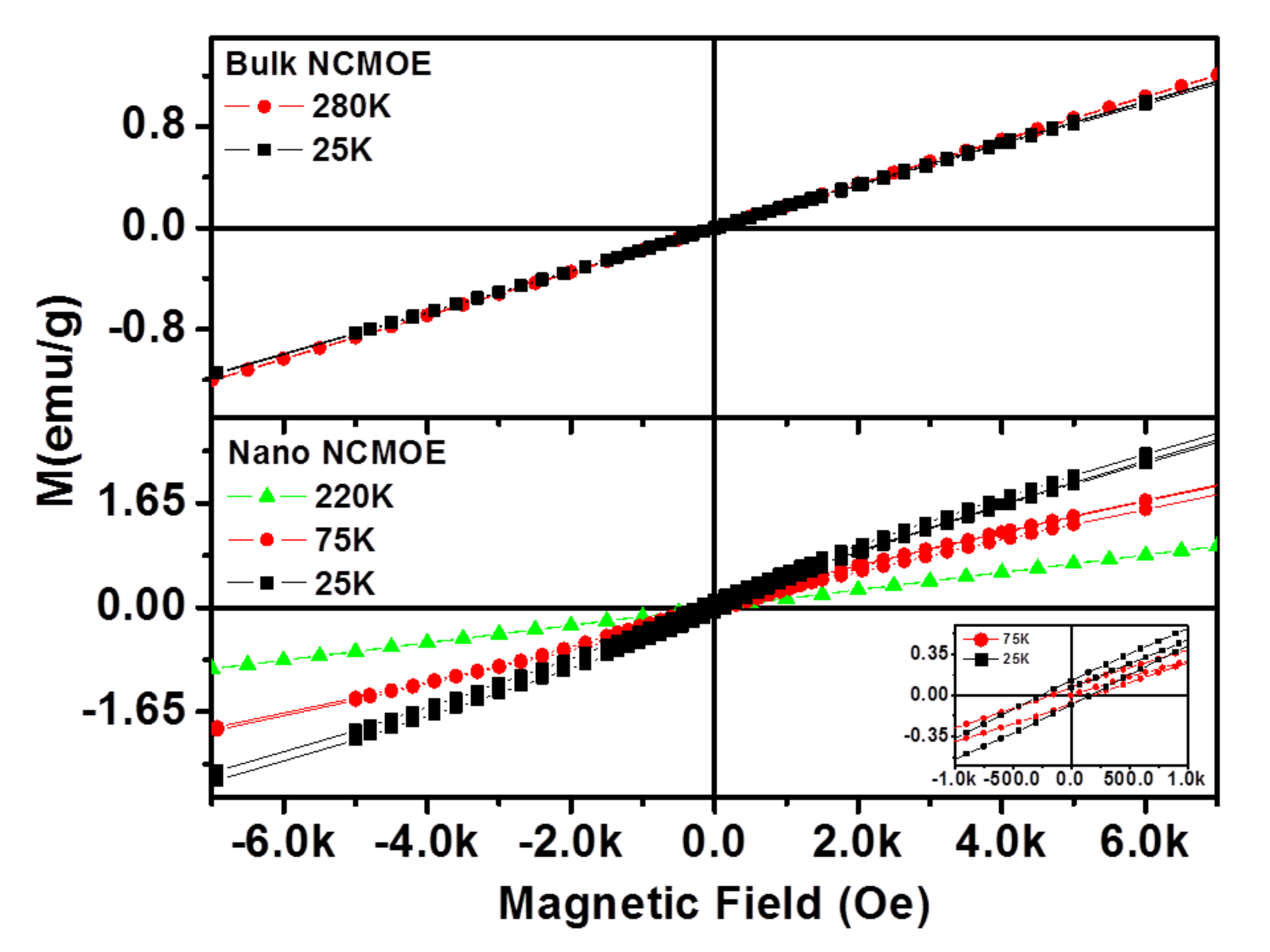}
\caption{ M vs H of NCMOE bulk and nano samples }
\label{MHe}
\end{figure}

\section{Electron Paramagnetic Resonance Studies}\label{EPR}
It is known that EPR is a sensitive probe of electron-hole asymmetry\cite{Joshi,Pad}. EPR spectra were recorded using a commercial X-band EPR spectrometer for all the four samples in the temperature range 4 to 300 K. Single, broad Lorentzian signals (in derivative form because of the magnetic field modulation and phase sensitive detection used) were observed at all but very low temperatures. A speck of DPPH was used as a field marker. The signals were fitted to the broad Lorentzian lineshape function\cite{Joshi1}

$\frac{dP}{dH} = A \frac{d}{dH} (\frac{\Delta H}{4(H-H_o)^2+(\Delta H)^2} + \frac{\Delta H}{4(H+H_o)^2+(\Delta H)^2})$

where P is the microwave power absorbed by the sample at resonance, $\Delta H$ is the linewidth, $A$ is a quantity proportional to the intensity of the signal and $H_o$ is the resonance field to extract the lineshape paramaters viz., intensity, resonance field $H_o$ and the linewidth. ‘g’ values were calculated using the relation $h\nu = g\beta H_o$ where $\nu$ is the microwave frequency, $h$ is the Planck constant and $\beta$ is the Bohr magneton. The temperature dependence of the ‘g’ parameters for the four samples is presented in Fig. 7 along with a few typical EPR signals in the inset.  It is seen that the ‘g’ values and their temperature dependences for the two bulk samples are quite different from each other. However, in the case of the nanosamples these differences decrease substantially and the behaviors of the two nanosamples closely approach each other. 

\begin{figure}
\includegraphics[width=\linewidth]{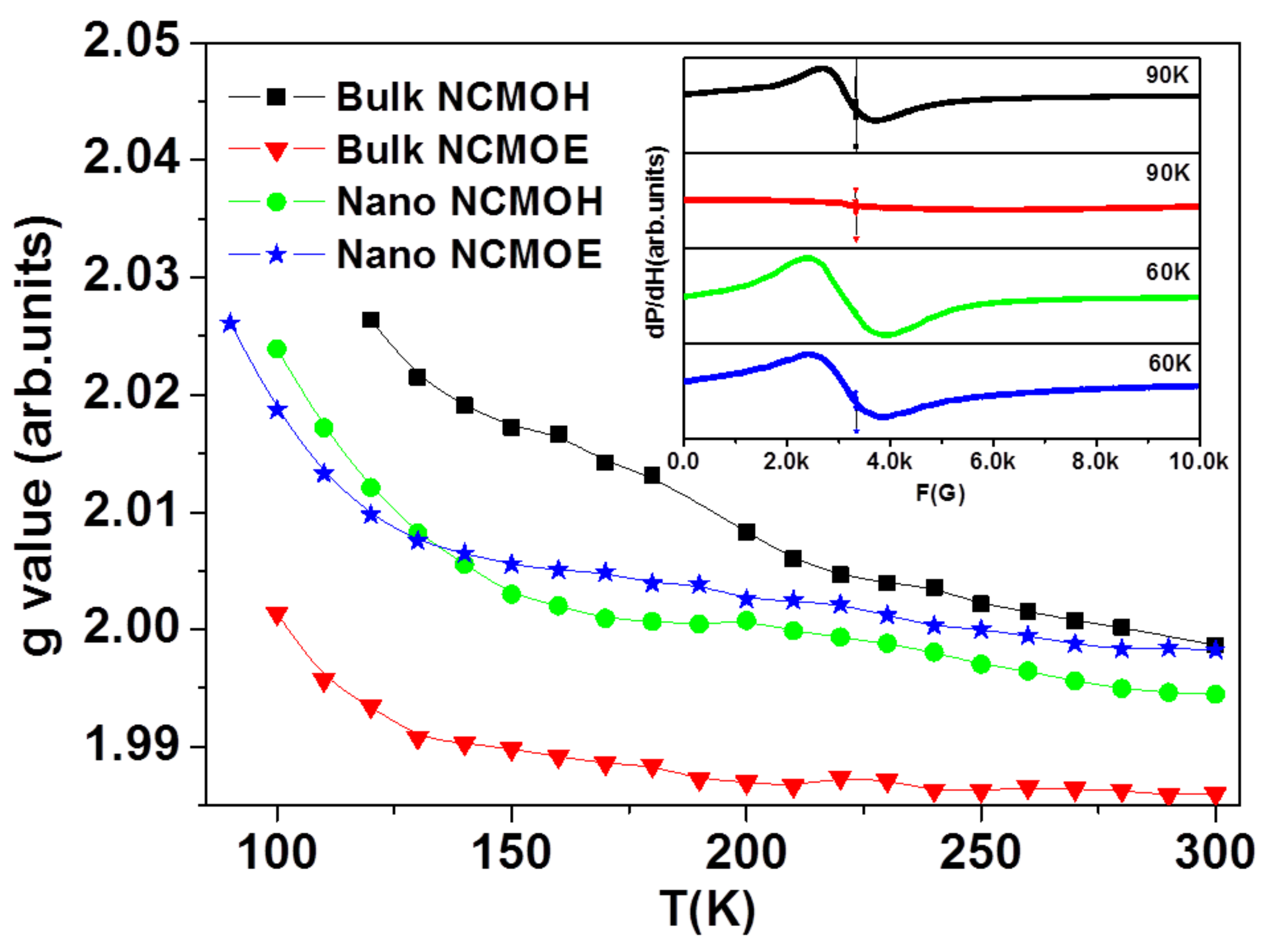}
\caption{ Temperature dependence of g-values of the samples; inset shows a selected EPR signal for each sample}
\label{MHe}
\end{figure}

\section{Discussion} \label{Discussion}
As mentioned earlier, our main interest in the present work is to study the effect of size reduction to nanoscale on the electron-hole asymmetry observed in the bulk samples. 
Magnetization studies clearly show the difference in the properties of the two bulk samples. The M(T) for bulk NCMOH shows a complex temperature dependence undergoing AFM and canted AFM/mixed phase transitions along with a CO transition. On the other hand bulk NCMOE undergoes only a CO transition and the increase in magnetization seen slightly below 25 K being due to the weak coupling of Nd ions with the Mn ions. 
When the size of the particles is reduced to $<100$ nm the properties change drastically leading to the disappearance of the number of phases present in the bulk samples and both the nanosamples exhibiting similar, single transition to a FM-like phase around the same transition temperature of $\sim100$ K resulting in the disappearance of electron-hole asymmetry. 
The EPR results further support this conclusion; the magnitudes and the temperature dependences of the 'g' parameters of bulk NCMOH and NCMOE are very different from each other. This difference is seen to considerably decrease when the size is reduced to nanodimensions.

While this experimental result is of fundamental importance, we can present only a qualitative explanation for it as even the phenomenon of the complexity of phase diagrams of the bulk manganites and the origin of the electron-hole asymmetry is poorly understood at present. It is believed that simultaneous existence of several competing interactions such as spin, charge, lattice and/or orbital in the presence of quenched disorder make many different phases equally probable and thus lead to the complexity of the manganite phase diagrams\cite{Dagotto1}.
We surmise that in the nanoparticles the absence of large length scales needed to fully capture the complex behavior of bulk manganites\cite{Dagotto1} leads to a collapse of the complexity.      

The disappearance of CO and occurrence of FM in nanomanganites in particular has been addressed in recent years~\cite{RaoNCMO,RaoPCMO,Singh} and is sought to be explained via the core-shell model~\cite{Zhang4,Zhang3,Dong1,Dong2}. According to this model, as the particle size is reduced to a few nanometers the surface to volume ratio increases considerably and the surface effects become prominent. The core of the nanoparticle retains its property with antiferromagnetic arrangement of the spins whereas the shell becomes FM due to the uncompensated surface spins. This results in different spin structures in the core and shell of the particle.  The magnetic behavior of these particles is strongly governed by the surface spins and the formation of the FM layer suppresses the CO in nanoparticles. However, more work, experimental as well as theoretical, is required to fully characterize and understand the 'simple' and near-universal phase diagram of nanomanganites.

\section{Conclusion} \label{Conclusion}
In conclusion, we have prepared the nanoparticles of NCMOH and NCMOE samples by citrate-gel method and the bulk samples by sintering the nanosamples at higher temperatures. These samples were characterized by XRD, EDAX, TEM, SQUID and EPR measurements. The two bulk samples show very different properties with NCMOH undergoing CO, AFM and canted AFM/mixed phase transitions whereas the NCMOE undergoes CO transition alone. In both the nanosamples, practically similar magnetic behavior is observed with the disappearance of most of the bulk phases and the emergence of an FM phase with  $T_c \sim 100$ K there by suppressing the electron-hole asymmetry seen in the bulk samples.

\small{\it{SVB thanks UGC and INSA for financial support and BKS thanks UGC for the meritorious fellowship.}}

\bibliography{main}

\end{document}